\documentstyle[12pt]{article}

\begin{document}

\input amssym.tex

\title{Approximative analytical solutions of the Dirac equation in 
Schwarzschild  spacetime}

\author{Ion I.  Cot\u aescu \thanks{E-mail:~~~cota@physics.uvt.ro}\\
{\small \it West University of Timi\c soara,}\\
       {\small \it V.  P\^ arvan Ave.  4, RO-300223 Timi\c soara, Romania}}


\maketitle

\begin{abstract}
Approximative analytic solutions of the Dirac equation in the
geometry of Schwarzschild black holes are derived obtaining
information about the discrete energy levels and the asymptotic
behavior of the energy eigenspinors.

 Pacs:
04.62.+v
\end{abstract}
\

The problem of the Dirac perturbations in the gravitational field of
a Schwarzschild black hole is less studied since the free Dirac
equation in this geometry is not analytically solvable at least in
the systems of coordinates and tetrad gauge fixings used until now.
However, despite of this difficulty, interesting results were
obtained using suitable analytical and numerical methods
\cite{S1}-\cite{S3}.

In this letter we would like to consider this problem  in a new
conjecture in which the local frames of the central static charts
are defined by the tetrad fields we have proposed some time ago
\cite{C1,ES}. This tetrad gauge shows off the central symmetry as a
global one such that the separation of the spherical variables can
be done as in special relativity \cite{TH}. After the separation of
the angular variables we remain with a pair of simple radial
equations depending only on the metric components. In this approach
we have found the energy eigenspinors of a massive Dirac field on
the de-Sitter \cite{C2} and anti-de Sitter spacetimes \cite{C1,C3}.
Unfortunately, this can not be achieved in the case of the
Schwarzschild background.

For this reason, our purpose is to find reasonable minor
approximations of the radial equations in which these could be
analytically solvable. In this way we obtain interesting information
concerning the discrete energy levels and the asymptotic behavior of
the energy eigenspinors. The results are presented in natural units
with $\hbar=c=1$.

In general, a manifold with central symmetry has a central static
chart with spherical coordinates $(t, r, \theta, \phi)$, covering
the space domain $D=D_{r}\times S^{2}$, i.e. $r\in D_{r}$ while
$\theta$ and $\phi$ cover the sphere $S^{2}$. For defining the local
frames we use the tetrad fields defined in Refs. \cite{C1,ES} whose
components depend on three arbitrary functions of $r$, denoted by
$u$, $v$ and $w$, which give the line element as
\begin{equation}\label{(muvw)}
ds^{2}=w^{2}\left[dt^{2}-\frac{dr^{2}}{u^2}-
\frac{r^2}{v^2}(d\theta^{2}+\sin^{2}\theta d\phi^{2})\right]\,.
\end{equation}

We have shown \cite{C1} that in these circumstances the separation
of the spherical variables of the free Dirac equation can be done as
in the case of the central problems in Minkowski flat spacetime.
Consequently, the Dirac field of mass $m$ can be written as a linear
combination of particular solutions of  given energy. Those of
positive frequency and energy $E$,
\begin{eqnarray}
U_{E,j,m_{j},\kappa_{j}}({x})&=&
U_{E,j,m_{j},\kappa_{j}}(t,r,\theta,\phi)\label{(u)}\\
&=&\frac{v(r)}{rw(r)^{3/2}}[f^{+}(r)\Phi^{+}_{m_{j},\kappa_{j}}(\theta,\phi)
+f^{-}(r)\Phi^{-}_{m_{j},\kappa_{j}}(\theta,\phi)]e^{-iEt}\,,\nonumber
\end{eqnarray}
are particle-like energy eigenspinors expressed in terms of radial
wave functions $f^{\pm}$ and  usual four-component angular spinors
$\Phi^{\pm}_{m_{j}, \kappa_{j}}$.  These spinors are orthogonal to
each other being completely determined by the angular quantum
numbers, $j$ and $m_{j}$, and the value of $\kappa_{j}=\pm (j+1/2)$
\cite{TH,BJD}. Moreover, they are normalized to unity with respect
to their own angular scalar product. We note that the
antiparticle-like energy eigenspinors  can be obtained directly
using the charge conjugation  as in the flat case \cite{C3}.

Thus the problem of the angular motion is completely solved. We
remain with  the radial wave functions $f^{\pm}$ which satisfy two
radial equations  that can be written in compact form as the
eigenvalue problem $H{\cal F}=E{\cal F}$ of the radial Hamiltonian
\cite{C1}
\begin{equation}\label{HR}
H=\begin{array}{|cc|}
    mw& -u\frac{\textstyle d}{\textstyle dr}+\kappa_{j}\frac{\textstyle v}
{\textstyle r}\\
&\\
  u\frac{\textstyle d}{\textstyle dr}+\kappa_{j}\frac{\textstyle v}
{\textstyle r}& -mw
\end{array}\,\,,
\end{equation}
in the space of  two-component vectors,  ${\cal F}=|f^{+},
f^{-}|^{T}$, equipped with the radial scalar product \cite{C1}
\begin{equation}\label{(spf)}
({\cal F}_{1},{\cal F}_{2})=\int_{D_{r}}\frac{dr}{u}\,
{\cal F}_{1}^{\dagger}{\cal F}_{2}\,.
\end{equation}
This selects the 'good' radial wave functions, i.e. square
integrable functions or tempered distributions, which enter in the
structure of the particle-like energy eigenspinors  (\ref{(u)}).

Let us consider now a Dirac particle of mass $m$ freely moving in
the gravitational field of a black hole of mass $M$ with  the
Schwarzschild line element
\begin{equation}\label{(le)}
ds^{2}=\left(1-\frac{r_{0}}{r}\right)dt^{2}-\frac{dr^{2}}{1-
\frac{\textstyle r_{0}}
{\textstyle r}}- r^{2} (d\theta^{2}+\sin^{2}\theta~d\phi^{2})\,,
\end{equation}
defined on the radial domain $D_{r}=(r_{0}, \infty)$ where
$r_0=2MG$. Hereby we identify the functions
\begin{equation}
u(r)=1-\frac{r_0}{r}\,, \quad v(r)=w(r)=\sqrt{1-\frac{r_0}{r}}\,,
\end{equation}
that give the radial Hamiltonian (\ref{HR}). The resulting radial
problem  can not be analytically solved as it stays. Therefore, if
we desire to avoid the numerical methods, we are forced to look for
suitable approximations of the radial equations.

In order to find an efficient method of approximation it is
convenient to introduce the new dimensionless variable
\begin{equation}
x=\sqrt{\frac{r}{r_{0}}-1}\,\in\,(0,\infty)
\end{equation}
and the notations
\begin{equation}
\mu=r_{0}m\,,\quad \epsilon=r_{0}E\,,
\end{equation}
which allow us to rewrite the radial problem as
\begin{equation}\label{rrr}
\begin{array}{|cc|}
    \mu\sqrt{1+x^{2}}-\epsilon\left(x+\frac{\textstyle 1}{\textstyle x}\right)
& -\frac{\textstyle 1}{\textstyle 2}\frac{\textstyle d}{\textstyle
dx}+
\frac{\textstyle \kappa_{j}}{\textstyle \sqrt{1+x^2}}\\
&\\
    \frac{\textstyle 1}{\textstyle 2}   \frac{\textstyle d}{\textstyle dx}+
\frac{\textstyle \kappa_{j}}{\textstyle \sqrt{1+x^2}} & -
    \mu\sqrt{1+x^{2}}-\epsilon\left(x+\frac{\textstyle 1}{\textstyle
x}\right)
\end{array}\,\,
\begin{array}{|c|}
f^{+}(x)\\
\\
\\
f^{-}(x)
\end{array}=0\,.
\end{equation}
Moreover, from Eq.(\ref{(spf)}) we find that the radial scalar
product in the new variable takes the form
\begin{equation}\label{(norm)}
({\cal F}_1,{\cal F}_2)=
2r_{0}\int_{0}^{\infty}dx\,\left(x+\frac{1}{x}\right) {\cal
F}^{\dagger}_1{\cal F}_2\,.
\end{equation}

{\em Near singularity}, for small values of $x\sim 0$, the exact
solutions can be approximated by ${\cal
F}_{s}=|f^{+}_{s},\,f^{-}_{s}|^{T}$ which satisfy the approximative
radial equations,
\begin{equation}
\begin{array}{|cc|}
    \mu-\frac{\textstyle \epsilon}{\textstyle x}
& -\frac{\textstyle 1}{\textstyle 2}\frac{\textstyle d}{\textstyle
dx}+
\kappa_{j}\\
&\\
    \frac{\textstyle 1}{\textstyle 2}   \frac{\textstyle d}{\textstyle dx}+\kappa_{j}
& -
    \mu-\frac{\textstyle \epsilon}{\textstyle x}
\end{array}\,\,
\begin{array}{|c|}
f^{+}_{s}(x)\\
\\
\\
f^{-}_{s}(x)
\end{array}=0\,,
\end{equation}
where we neglected the terms of the order $O(x)$. Performing the
transformation ${\cal F}_s\to \hat{\cal F}_{s}=U{\cal F}_{s}=|\hat
f^{+}_{s},\,\hat f^{-}_{s}|^T$, with the help of the unitary matrix
\begin{equation}
U=\frac{1}{\sqrt{2}}\left|
\begin{array}{cc}
1&i\\
i&1
\end{array}\right|\,,
\end{equation}
we obtain the simpler system of equations
\begin{equation}
\left(\frac{1}{2}\frac{d}{dx}\pm\frac{i\epsilon}{x}\right) \hat
f^{\pm}_{s}(x)=(\mu\pm i\kappa_{j})\hat f^{\mp}_{s}(x)\,.
\end{equation}
This can be analytically solved in terms of Bessel functions $J_l$
and $Y_l$ as \cite{AS}
\begin{eqnarray}
\hat f_s^+(x)&=&\sqrt{\kappa_j -i\mu}\left(c_1
\sqrt{x}\,J_{l_+}(2i\lambda_j x)+c_2\sqrt{x}\,
Y_{l_+}(2i\lambda_j x)\right)\,,\\
\hat f_s^-(x)&=&\sqrt{\kappa_j+i\mu}\left(c_1
\sqrt{x}\,J_{l_-}(2i\lambda_j x)+c_2\sqrt{x}\, Y_{l_-}(2i\lambda_j
x)\right)\,,
\end{eqnarray}
where $c_1$ and $c_2$ are arbitrary constants and
\begin{equation}
l_{\pm}=2i\epsilon\pm \frac{1}{2}\,,\quad
\lambda_{j}=\sqrt{\kappa_{j}^2+\mu^2}\,.
\end{equation}
We observe that near singularity these wave functions are regular
but their phases are not determined since  $\hat f_s^{\pm}\propto
x^{\mp 2i\epsilon}+O(x) $ for $x\to 0$.

{\em The asymptotic behavior}, for very large values of $x$, can be
studied approximating $\sqrt{1+x^2}\sim x$ and obtaining thus an
asymptotic radial problem for ${\cal
F}_{a}=|f^{+}_{a},\,f^{-}_{a}|^{T}$. The asymptotic radial equations
can be put in the form
\begin{equation}\label{RA}
\begin{array}{|cc|}
 \frac{\textstyle 1}{\textstyle 2} \frac{\textstyle d}{\textstyle dx}
 +\frac{\textstyle\kappa_j}{\textstyle x}
& -\mu\left( x+\frac{\textstyle\delta}{\textstyle x}\right)
-\epsilon \left(x+\frac{\textstyle 1}{\textstyle x}\right)\\
&\\
 -\mu \left(x+\frac{\textstyle\delta}{\textstyle x}\right)
 +\epsilon\left(x+\frac{\textstyle 1}{\textstyle x}\right)
&
   \frac{\textstyle 1}{\textstyle 2} \frac{\textstyle d}{\textstyle dx}
   -\frac{\textstyle \kappa_j}{\textstyle x}
\end{array}\,\,
\begin{array}{|c|}
f^{+}_{a}(x)\\
\\
\\
f^{-}_{a}(x)
\end{array}=0\,,
\end{equation}
where we introduced the new parameter $\delta\in [0,1)$ which could
play the role of a fit parameter. Of course, if we rigorously
consider the approximation of the order $O(1/x)$ then we must take
$\delta=\frac{1}{2}$.

Since Eqs. (\ref{RA}) can be solved for any value of $\epsilon$, we
have to look separately for solutions corresponding either to a
discrete energy spectrum when $\epsilon<\mu$ or to a continuous one
in the domain $[\mu, \infty)$. For finding these solutions,  we
diagonalize the term proportional to $x$ of the operator (\ref{RA})
using the transformation matrix
\begin{equation}
T=\frac{1}{2\nu}\left|
\begin{array}{cc}
\sqrt{\mu-\epsilon}&\sqrt{\mu+\epsilon}\\
-\sqrt{\mu-\epsilon}&\sqrt{\mu+\epsilon}\\
\end{array}\right|\,.
\end{equation}
where we denote $\nu=\sqrt{\mu^2-\epsilon^2}$. After the
transformation ${\cal F}_a\to \hat{\cal F}_{a}=T{\cal F}_{a}= |\hat
f^{+}_{a},\,\hat f^{-}_{a}|^T$ we obtain the new equations
\begin{equation}\label{TE}
\left[\frac{1}{2} x
\frac{d}{dx}\pm\left(\frac{\epsilon^2-\delta\mu^2}{\nu}-\nu x^2
\right)\right]\hat f^{\pm}_a
=\left(\kappa\mp\frac{\epsilon\mu(\delta-1)}{\nu}\right)\hat
f^{\mp}_a\,,
\end{equation}
that can be analytically solved.

{\em The discrete quantum modes}  with discrete energy levels
$\epsilon<\mu$ are governed by radial wave functions which must be
square integrable. After a little calculation we find that Eqs.
(\ref{TE}) allow the particular solutions
\begin{eqnarray}
\hat f^+_a(x)&=&N^+ x^{2s_j} e^{-\nu x^2} {_1F_1}(s_j+q+1,2s_j+1,2\nu x^2)\,,\label{E1}\\
\hat f^-_a(x)&=&N^- x^{2s_j} e^{-\nu x^2}
{_1F_1}(s_j+q\,,2s_j+1,2\nu x^2)\,,\label{E2}
\end{eqnarray}
where the confluent hypergeometric functions $_1F_1$ \cite{AS}
depend on the following real parameters
\begin{equation}
s_j=\sqrt{\kappa_j^2+\nu^2+\mu^2 (\delta^2-1)}\,, \quad q=\nu +
\frac{\mu^2(\delta-1)}{\nu}\,.
\end{equation}
These solutions satisfy Eqs. (\ref{TE}) only if the normalization
constants $N^{\pm}$ obey
\begin{equation}
\frac{N^-}{N^+}=\frac{\nu(s_j-q)}
{\kappa_j\nu-\mu\epsilon(\delta-1)}\,.
\end{equation}
We obtain thus the transformed radial wave functions (\ref{E1}) and
(\ref{E2}) that have a good asymptotic behavior and are regular in
$x=0$.

These solutions may be square integrable with respect to the scalar
product (\ref{(norm)}) only when we impose the quantization
condition $s_j+q=-n\,,\, n=1,2,3,...$. This leads to the equation
\begin{equation}\label{quant}
\sqrt{\kappa_j^2+\nu^2+\mu^2 (\delta^2-1)}+\nu +
\frac{\mu^2(\delta-1)}{\nu}=-n\,,
\end{equation}
which is of the third order in $\nu$ and can be solved at anytime
under Mathematica or Maple, obtaining thus the {\em asymptotic}
energy spectrum. However, the analytic formula of these energy
levels is too complicated to be written here. For this reason, we
restrict ourselves to give some reasonable estimations. First we
observe that the energies levels, $E_{n,j}$, must  satisfy the
conditions
\begin{equation}
\frac{\kappa_j^2}{\kappa_j^2+\mu^2(1-\delta)^2}\le\frac{{E_{n,j}}^2}{m^2}<1\le
\delta^2+\frac{\kappa_j^2}{\mu^2}\,.
\end{equation}
The last inequality prevents the Dirac particles to get near to the
singularity where the black hole could capture them. For $\mu<1$ and
small values of $\nu$ we have to use the approximative formula
\begin{equation}
E_{n,j}\simeq
m\left[1-\frac{\mu^2(1-\delta)^2}{\left(n+\sqrt{\kappa_j^2
+\mu^2(\delta^2-1)}\right)^2}\right]^{\frac{1}{2}}\,.
\end{equation}
Moreover, in the case of $\mu\ll 1$  and $\delta =\frac{1}{2}$ we
recover the Newtonian result
\begin{equation}
E_{n,j}\sim m\left(
1-\frac{1}{8}\frac{\mu^2}{(n+j+\frac{1}{2})^2}\right)
=m-\frac{1-e^2}{2}\frac{G^2M^2m^3}{L^2}\,.
\end{equation}
for a particle of mass $m=\mu/r_0$ and angular momentum $L=j\pm
\frac{1}{2}$ moving on an ellipsoidal  trajectory of eccentricity
\begin{equation}
e=\left(1-\frac{(j\pm\frac{1}{2})^2}{(n+j+\frac{1}{2})^2}\right)^{\frac{1}{2}}\,.
\end{equation}

These asymptotical results, including the exact solutions of Eq.
(\ref{quant}), must be considered prudently trying to verify and
improve them using numerical methods.   We note that only the
massive fermions can be retained in Schwarzschild gravitational
fields on discrete levels while the massless fermions can not do
this.

{\em The scattering modes} corresponding to the continuous energy
spectrum $\epsilon\in [\mu,\infty)$ are described by particular
solutions that behave as tempered distributions. In this case we
introduce the real parameter $\hat\nu=\sqrt{\epsilon^2-\mu^2}$
instead of $\nu$. Following the same procedure like in the previous
case we look for the solutions of the transformed equations
(\ref{TE}) where we replace $\nu=i\hat\nu$. Hereby we obtain the
transformed radial wave functions in terms of Whittaker functions
$M_{p,s}$ and $W_{p,s}$ as \cite{AS}
\begin{eqnarray}
\hat f^+_a(x)&=&C_1^+\frac{1}{x}M_{p_+,s_j}(2i\hat\nu x^2)
+C_2^+\frac{1}{x}W_{p_+,s_j}(2i\hat\nu x^2)\label{E11}\\
\hat f^-_a(x)&=&C_1^-\frac{1}{x}M_{p_-,s_j}(2i\hat\nu x^2)
+C_2^-\frac{1}{x}W_{p_-,s_j}(2i\hat\nu x^2)\,,\label{E22}
\end{eqnarray}
where we denote $p_{\pm}=\mp\frac{1}{2}-q $ while the constants
satisfy
\begin{equation}
\frac{C_1^-}{C_1^+}=\frac{\hat\nu(s_j-q)}{\kappa_j\hat\nu+i\epsilon\mu
(\delta-1)}\,,\quad
\frac{C_2^-}{C_2^+}=-\frac{\hat\nu}{\kappa_j\hat\nu+i\epsilon\mu
(\delta-1)}\,.
\end{equation}
These solutions are useful for analyzing the scattering of the Dirac
particles on black holes.

Hence we have all the elements we need to write down the
approximative form of the energy eigenspinors  in all of the above
discussed cases. Starting with the approximative form of the
transformed radial wave functions $\hat f^{\pm}$ we have to
calculate the functions $f^{\pm}$ which should give the
particle-like energy eigenspinors (\ref{(u)}). The antiparticle-like
energy eigenspinors can be calculated using the charge conjugation
as \cite{C3}
\begin{equation}
V_{E,j,m_{j},\kappa_j}=(U_{E,j,m_{j},\kappa_j})^{c}\equiv C
(\overline{U}_{E,j,m_{j},\kappa_j})^{T} \,,\quad
C=i\gamma^{2}\gamma^{0}\,.
\end{equation}

We hope that the approximative solutions presented here should help
one to improve the analytical or numerical methods for studying the
Dirac field in the Schwarzschild background.

\end{document}